\documentclass{article}

\newcommand{\p}[1]{(\ref{#1})}

\newcommand{\bF}{{\overline F}}

\newcommand{\bX}{{\overline X}}

\newcommand{\bD}{{\overline D}}

\newcommand{\bW}{{\overline W}}

\newcommand{\cW}{{\cal W}}
\newcommand{\cX}{{\cal X}}
\newcommand{\cM}{{\cal M}}

\newcommand{\be}{\begin{equation}}
\newcommand{\ee}{\end{equation}}
\newcommand{\bea}{\begin{eqnarray}}
\newcommand{\eea}{\end{eqnarray}}

\newcommand{\ba}{\begin{array}} \newcommand{\ea}{\end{array}}

\def\im{{\rm i}}

\newcommand{\nn}{\nonumber}

\usepackage{amscd,amsmath,amssymb}

\begin{document}
\thispagestyle{empty}
\vspace{2cm}
\begin{flushright}
\end{flushright}\vspace{2cm}
\begin{center}
{\Large\bf Comments on N=2 Born-Infeld Attractors}
\end{center}
\vspace{1cm}

\begin{center}
{\large\bf S.~Bellucci${}^a$, S.~Krivonos${}^{b}$,
A.~Sutulin${}^{a,b}$}
\end{center}

\begin{center}
${}^a$ {\it
INFN-Laboratori Nazionali di Frascati,
Via E. Fermi 40, 00044 Frascati, Italy} \vspace{0.2cm}

${}^b$ {\it
Bogoliubov  Laboratory of Theoretical Physics, JINR,
141980 Dubna, Russia} \vspace{0.2cm}

\end{center}
\vspace{2cm}

\begin{abstract}\noindent
We  demonstrated that the new $N=2$ Born-Infeld action with two $N=1$ vector supermultiplets, i.e. $n=2$ case considered as the example in the recent paper by S.~Ferrara, M.~Porrati and A.~Sagnotti \cite{FPS}, is
 some sort of  complexification of J.~Bagger and A.~Galperin construction of $N=2$ Born-Infeld action \cite{BG}. Thus,  novel features
could be expected only for $n>2$ cases,  if the standard action is
considered.

\end{abstract}

\newpage
\setcounter{page}{1}
\setcounter{equation}{0}
\section{Introduction}
In the  recent interesting paper \cite{FPS} S.~Ferrara, M.~Porrati and A.~Sagnotti proposed the generalization of J.~Bagger and A.~Galperin construction of $N=2$ Born-Infeld action \cite{BG} to the cases of several $N=1$ vector supermultiplets. Roughly speaking, they proposed to generalize the non-linear constraint of \cite{BG}
\be\label{eq1}
W\cdot W +X \left( m -\frac{1}{4} \bD{}^2 \bX \right) =0
\ee
to the case of many vector multiplets as
\be\label{eq2}
d_{abc} \left( W_a \cdot W_b +X_a \left( m_b -\frac{1}{4} \bD{}^2 \bX_b \right)\right) =0,
\ee
where $d_{abc}$ are totally symmetric constants, $m_a$ is a set of constants and {\it dot} means converting of the Lorentz indices. The invariance of the constraint \p{eq2} with respect to the hidden $N=1$ supersymmetry transformations
\be\label{eq3}
\delta \left( W_a\right)_\alpha = \left( m_a -\frac{1}{4} \bD{}^2 \bX_a \right)\eta_\alpha -\im \partial_{\alpha\dot\alpha} X_a \bar\eta{}^{\dot\alpha}, \;
\delta X_a = -2 \left( W_a\right)^\alpha \eta_\alpha ,
\ee
introduced the additional constraint
\be\label{eq4}
d_{abc} \left(W_b\right)^\alpha \; X_c =0 .
\ee
In the case of one supermultiplet $(n=1)$ the constraint \p{eq4} trivially follows from \p{eq1}, but if $n >1$ it has to be additionally taken into account. The constraints \p{eq2}, \p{eq4},
being solved, expressed the bosonic $N=1$ chiral superfields $X_a$ in terms of the $N=1$ vector supermultiplets $(W_a)_\alpha$. In virtue of the transformation properties \p{eq3} and the property $D^2 W_\alpha \sim \partial_{\alpha\dot\alpha} \bW^{\dot\alpha}$,
the trivial action which is invariant with respect to the hidden $N=1$ supersymmetry
\be\label{eq5}
S= \int d^4x d^2 \theta \; e^a X_a + c.c., \qquad  e^a = const,
\ee
becomes meaningful and its bosonic core in the case $n=1$ is just the Born-Infeld action \cite{BG}.

In the paper \cite{FPS} the Authors presented the detailed analysis of $n=2$ case, which can be divided into two subcases:
\begin{itemize}
\item $d_{111}=1, d_{112}=-1 \quad I_4=0$,
\item $d_{111}=1, d_{122}=-1 \quad I_4>0$
\end{itemize}
where $I_4$ is a quartic invariant, discussed in \cite{FPS}.

Leaving aside the case with $I_4=0$, which results in the system
without kinetic term for one of the $N=1$ vector supermultiplet
(with the action \p{eq5}), let us consider in  details the second
case.

\setcounter{equation}{0}
\section{$n=2, I_4>0$ case}
With the choice $d_{111}=1, d_{122}=-1$ the constraints \p{eq2}, \p{eq4} read
\bea\label{meq}
&& \left[ W_1 \cdot W_1 +X_1 \left( m_1 -\frac{1}{4} \bD{}^2 \bX_1 \right)\right]-\left[ W_2 \cdot W_2 +X_2 \left( m_2 -\frac{1}{4} \bD{}^2 \bX_2 \right)\right] =0, \nn \\
&&2 W_1 \cdot W_2 +X_1 \left( m_2 -\frac{1}{4} \bD{}^2 \bX_2 \right)+X_2 \left( m_1 -\frac{1}{4} \bD{}^2 \bX_1 \right)=0, \nn \\
&& \left(W_1\right)^\alpha \; X_1 -\left(W_2\right)^\alpha \; X_2 =0,\qquad \left(W_1\right)^\alpha \; X_2 +\left(W_2\right)^\alpha \; X_1 =0 .
\eea
If we now introduce the new superfields and constants
\bea\label{newW}
&&\cW_1 = W_1+\im W_2,\; \cW_2 = W_1-\im W_2,\quad \widetilde \cW_1 =\bW_1+\im \bW_2, \; \widetilde \cW_2 =\bW_1-\im \bW_2, \nn \\
&&\cX_1 = X_1+\im X_2,\; \cX_2 = X_1-\im X_2,\quad \widetilde \cX_1 =\bX_1+\im \bX_2, \; \widetilde \cX_2 =\bX_1-\im \bX_2, \nn \\
&&\cM_1 = m_1+\im m_2,\; \cM_2 = m_1-\im m_2,\quad \widetilde \cM_1 =\bar{m}_1+\im \bar{m}_2, \; \widetilde \cM_2 =\bar{m}_1-\im \bar{m}_2,
\eea
then the system of the equations \p{meq} acquires a ``decoupled'' form to become
\be\label{meq1}
 \cW_a \cdot \cW_a +\cX_a \left( \cM_a -\frac{1}{4} \bD{}^2 \widetilde\cX_a \right)=0, \quad \left(\cW_a\right)^\alpha \; \cX_a=0.\qquad{\mbox{\bf no summation over "a"!}}
\ee
Clearly, the solution for each of the equations \p{meq1} will be the same as in \cite{BG} with the replacement
\be
\left\{W, \bW, X, \bX\right\} \quad \Rightarrow \quad \left\{ \cW_a, \widetilde\cW_a, \cX_a, \widetilde\cX_a\right\}.
\ee
Thus, we are dealing with the some sort of  complexification ("tilde conjugation") of the Born-Infeld theory action. On the bosonic level this yields the following action:
\be\label{bos}
S_{bos} = S_{BI} \Big[(F_1)_{\alpha\beta}+\im (F_2)_{\alpha\beta}, (\bF_1)_{\dot\alpha \dot\beta}+\im (\bF_2)_{\dot\alpha \dot\beta}\Big] +c.c.
\ee

The origin of such decoupling may be understood as follows. If we introduce the matrices $d_1$ and $d_2$ constructed from the constants $d_{abc}$, i.e.
\be\label{mat1}
d_1 = \left( \begin{array}{cc}
             d_{111} & d_{112}\\
             d_{121} & d_{122}
             \end{array} \right) = \left( \begin{array}{cc}
             1 & 0\\
             0 & -1
             \end{array} \right), \quad
             d_2 = \left( \begin{array}{cc}
             d_{211} & d_{212}\\
             d_{221} & d_{222}
             \end{array} \right) = \left( \begin{array}{cc}
             0 & -1\\
             -1 & 0
             \end{array} \right),
\ee
then the decoupling means that the matrix $ D=\alpha\; d_1 + \beta \; d_2$ has coinciding eigenvalues for two different sets of  coefficients $\alpha, \beta$.
In the present case the characteristic equation for the matrix $D$ reads
\be
\lambda^2 -\alpha^2 -\beta^2 =0,
\ee
and, therefore, the decoupling is possible and $ \alpha = \pm \im \beta$.

It should be noted that in the ``diagonal basis'' \p{meq1}
$\;(d_{111}=1, d_{222}=1) \; \rightarrow \; I_4 <0$. Thus, the
change of variables \p{newW} maps the $I_4>0$ case into the
$I_4<0$ one.

Therefore, the case with two vector supermultiplets is not instructive, being just  a special complexification of the
Born-Infeld action for the complex vector multiplet. One may expect that the novel features would appear in the next case with $(n=3)$, where the ''diagonalization'' of the constraints \p{eq2} and \p{eq4} is not obvious.

Let us note, that the proposed generalization of the constraints will work perfectly for many other systems with partial  breaking of  supersymmetry, for which
the description {\it \`a la} Bagger and Galperin is known. Among such systems there are the supermembrane in $D=4$ \cite{IK1}, the L3 brane on $AdS_5$ \cite{BIK1},  superparticles \cite{DIK1}, etc.
So, if the next $n=3$ case gives rise to a new non-trivial system, than it would be very interesting to analyze above mentioned systems too.

Finally, we would like to mention that the situation with this
generalization is quite similar to the rational Calogero model
(and its supersymmetrization)  where the two particles case is
almost trivial, while the first interesting system starts from the
three particles one. \setcounter{equation}{0}\vspace{0.5cm}

\noindent{\large Note added.} All that we said above is correct
for the action \p{eq5} only. If in the action  the term with
$C_{AB}$ is added, as it was done in \cite{FPS}, the situation
changes drastically. From the eq.(1.16) in the paper \cite{FPS}
one may learn that in the $n=2$ case this causes the appearance of
$\int d^4 \theta Y^A {\bar Y^B}$ terms in the action. Such terms
could produce higher derivative fermionic terms in the component
level action. It is not immediately clear that such new terms will
not result in non-standard equations of motion for the fermions,
as it happened in the $n=1$ case \cite{BG}. In our opinion, this
point has to be further clarified.\vspace{0.5cm}

\section*{Acknowledgments}
We acknowledge correspondence with S.~Ferrara, A.~Sagnotti and
M.~Porrati concerning the additional $C_{AB}$ term in their
action. This work was partially supported by  the ERC Advanced
Grant no. 226455 \textit{``Supersymmetry, Quantum Gravity and
Gauge Fields''}~(\textit{SUPER\-FIELDS}). The work of S.K. was
supported by RSCF grant 14-11-00598.


\begin{thebibliography}{99}
\bibitem{FPS} S.~Ferrara, M.~Porrati, A.~Sagnotti, {\it N=2 Born Infeld Attractors}, {\tt arXiv:1411.4954 [hep-th]}.
\bibitem{BG} J.~Bagger, A.~Galperin, Phys. Rev. D {\bf 55} (1997) 1091, {\tt arXiv:hepth/9608177}.
\bibitem{IK1} E.~Ivanov, S.~Krivonos, Phys.Lett. {\bf B453} (1999) 237; Erratum-ibid. {\bf B657} (2007) 269, {\tt  arXiv:hep-th/9901003}.
\bibitem{BIK1} S.~Bellucci, E.~Ivanov, S.~Krivonos, Nucl.Phys. {\bf B672} (2003) 123, {\tt  arXiv:hep-th/0212295}.
\bibitem{DIK1} F.~Delduc, E.~Ivanov, S.~Krivonos, Nucl.Phys. {\bf B576} (2000) 196, {\tt arXiv:hep-th/9912222}.
\end{thebibliography}
\end{document}